\documentclass[11pt,a4paper]{article}
\pdfoutput=1
\usepackage{jheppub}
\makeatletter
\renewcommand{\@fpheader}{Preprint}
\makeatother
\usepackage{booktabs}
\usepackage{array}
\usepackage{placeins}
\newcommand{\piz}{\pi^0}
\newcommand{\metric}{\mathcal{M}}
\newcommand{\nnh}{\nu\bar{\nu}H}
\newcommand{\qqh}{q\bar qH}

\newcommand{\eventdisplay}[1]{\includegraphics[width=\linewidth,height=0.60\textheight,keepaspectratio]{#1}}

\title{Machine-learning-assisted photon pairing for neutral-pion reconstruction at future Higgs factories}
\author[a,b]{Yi Wang,}
\author[a,b]{Chao Yang,}
\author[a,b]{Tianyi Yang,}
\author[a,b]{Enrui Tian,}
\author[a,d]{Yuexin Wang,}
\author[c]{\newline Song Pan,}
\author[c]{Jun Guo,}
\author[a]{and Manqi Ruan}
\affiliation[a]{Institute of High Energy Physics, Chinese Academy of Sciences,\\
Beijing 100049, China}
\affiliation[b]{University of Chinese Academy of Sciences,\\
Beijing 100049, China}
\affiliation[c]{Shanghai Jiao Tong University,\\
Shanghai 200000, China}
\affiliation[d]{China Spallation Neutron Source Science Center, Dongguan 523803, China}
\emailAdd{wangyi95@ihep.ac.cn}
\emailAdd{yangchao@ihep.ac.cn}
\emailAdd{ruanmq@ihep.ac.cn}
\emailAdd{jun.guo@sjtu.edu.cn}

\abstract{
Neutral-pion ($\piz$) reconstruction is important for physics measurements at high-energy colliders, particularly in flavor, $\tau$, and Higgs physics. We develop four machine-learning-based methods for $\piz$ candidate selection: a boosted decision tree (BDT), a multilayer perceptron (MLP), ParticleNet (PN), and Particle Transformer (ParT). These methods are compared with the classical baseline (MW), which uses an optimized diphoton mass window, in four benchmark physics processes at electron--positron colliders. Performance is evaluated using the product of selection efficiency and purity. At the reference ECAL energy resolution, all four machine-learning methods outperform the classical baseline. PN and ParT achieve relative improvements of $53\%$--$95\%$ and $60\%$--$128\%$, respectively, primarily through improved rejection of combinatorial background. All four methods maintain their performance advantage across the investigated range of ECAL energy resolutions, which covers representative values for experiments at the high-energy frontier.
}

\begin{document}
\maketitle
\raggedbottom

\section{Introduction}
\label{sec:intro}

Neutral-pion ($\piz$) reconstruction is critical for physics measurements, especially in flavor physics. It is highly relevant to future Higgs and $Z$ factories, such as the Circular Electron Positron Collider (CEPC) and the electron--positron stage of the Future Circular Collider (FCC-ee), operating at center-of-mass energies of $\sqrt{s}\simeq91.2~\mathrm{GeV}$ and $240~\mathrm{GeV}$~\cite{CEPC_CDR_Physics,FCCee_CDR}. In particular, accurate $\piz$ reconstruction is essential for identifying hadronic $\tau$ decay modes~\cite{ATLAS:tauPFlow}, which is important for precision determinations of the strong coupling constant~\cite{PDG2024} and measurements of the Higgs boson CP properties in $H\to\tau^+\tau^-$ decays~\cite{Jeans:2018anq}.

$\piz$ has a branching fraction of $98.8\%$ for the decay into two photons~\cite{PDG2024}. The primary task of $\piz$ reconstruction is therefore to identify the corresponding photon pairs. A conventional approach is to select photon pairs with invariant masses close to the nominal $\piz$ mass, as commonly employed at Belle, Belle II, and BESIII~\cite{Belle:tauPiPi0,BelleII:B0pi0pi0,BESIII:chic4pi0}. The major limitation of this mass-window method is the combinatorial background, which becomes more severe in events with high photon multiplicities.

We develop four machine-learning methods for $\piz$ reconstruction, using a boosted decision tree (BDT) implemented with XGBoost~\cite{Chen:2016btl}, a multilayer perceptron (MLP), ParticleNet (PN)~\cite{ParticleNet}, and Particle Transformer (ParT)~\cite{Qu:2022mxj}. These methods utilize information from the full collection of reconstructed photons in each event~\cite{Thomson:2009rp,Ruan:2018yrh,Wemmer:2023}. Candidate photon pairs are first identified using a loose diphoton mass-window cut to improve computational efficiency. Information from the identified photon pairs and the remaining photons is then fed into the models in different representations. The models are trained via supervised learning to determine whether a candidate pair originates from a $\piz$ decay. These methods are compared with a classical baseline method using an optimized diphoton mass interval, denoted as the classical baseline (MW) throughout this paper. 

We quantify and compare the performance of the four machine-learning methods and the classical baseline (MW) in four benchmark physics processes, and investigate their dependence on the ECAL energy resolution and the training-sample size. Section~\ref{sec:methods} introduces the event samples, candidate-selection methods, and performance measures. Section~\ref{sec:results} presents and discusses the results, and section~\ref{sec:summary} summarizes the main findings. Additional performance results are provided in appendix~\ref{sec:detailed_results}, while all technical details are summarized in appendix~\ref{sec:technical_details}.

\section{Samples, selections and performance measure}
\label{sec:methods}

\subsection{Event samples}
\label{sec:samples}

\begin{figure}[tbp]
\centering
\includegraphics[width=\textwidth]{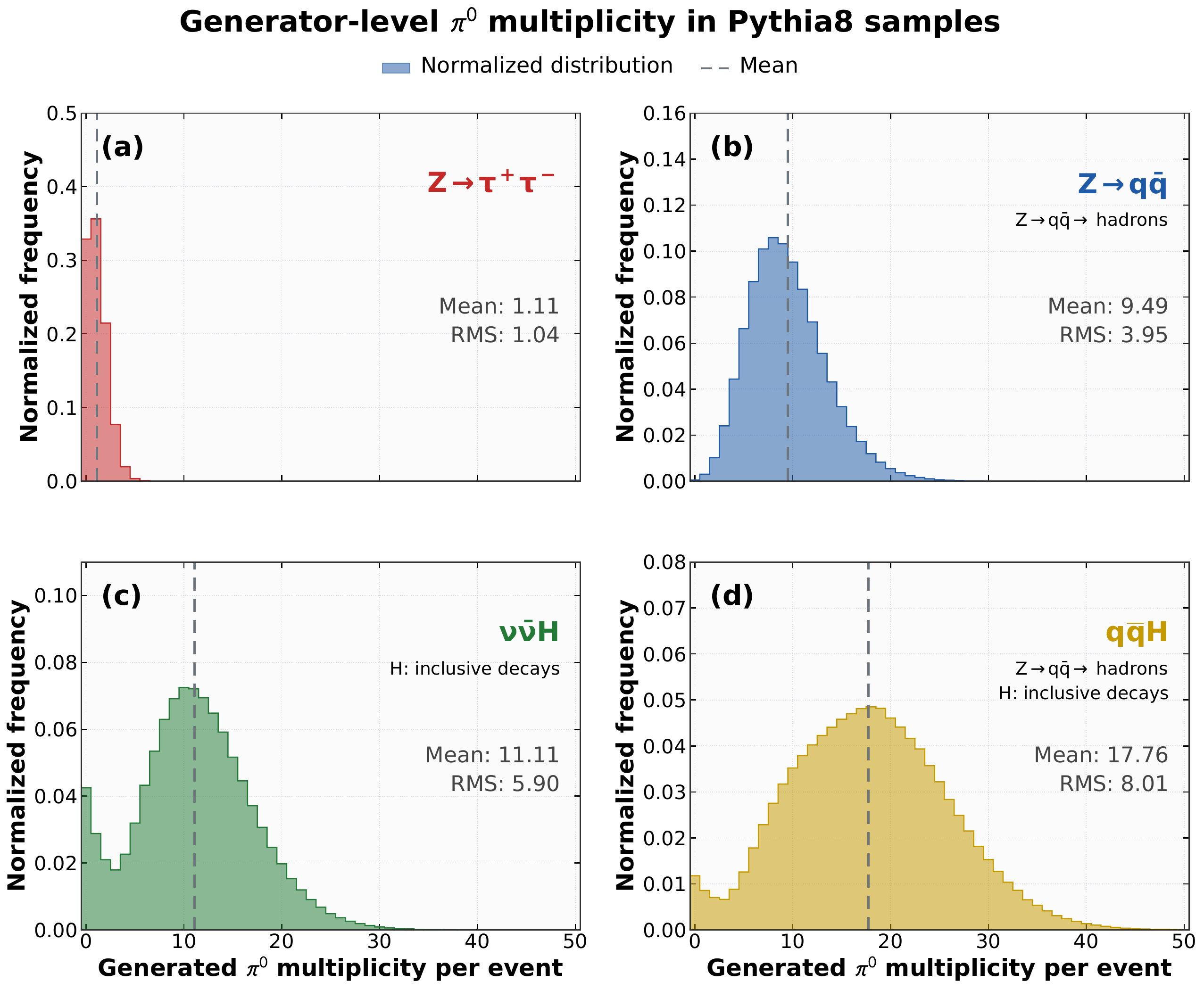}
\caption{Generated $\piz$ multiplicity for (a) $Z\to\tau^+\tau^-$, (b) $Z\to q\bar q$, (c) $\nnh$ and (d) $\qqh$. The Higgs decays are inclusive. Distributions are normalized to unit area.}
\label{fig:multiplicity}
\end{figure}

We consider four representative physics processes at future electron--positron colliders: $Z\to\tau^+\tau^-$ and $Z\to q\bar q$ at $\sqrt{s}=91.2~\mathrm{GeV}$, and $\nnh$ and $\qqh$ at $\sqrt{s}=240~\mathrm{GeV}$. The first two processes represent leptonic and hadronic $Z$ decays at the $Z$ pole. The $\nnh$ process includes Higgsstrahlung with $Z$ decaying into neutrinos, $WW$ fusion, and their interference, while the $\qqh$ sample consists of Higgsstrahlung events with the $Z$ boson decaying hadronically. The Higgs boson decays inclusively in both samples. These four processes are key physics processes at future $Z$ and Higgs factories, and their different photon multiplicities allow us to characterize the performance of $\piz$ reconstruction comprehensively. PYTHIA 8~\cite{Bierlich:2022pfr} is used for Monte Carlo event generation, yielding average $\piz$ multiplicities of 1.11, 9.49, 11.11, and 17.76 per event for the four benchmark processes, respectively (figure~\ref{fig:multiplicity}). The detector response is simulated with Delphes 3.5.0~\cite{deFavereau:2013fsa}, using a detector card developed for CEPC studies~\cite{CEPC4thDelphesCard,CEPCFlavorPhysics}, featuring a homogeneous crystal ECAL as adopted in the CEPC Reference Detector~\cite{CEPC_RefTDR}.

The electromagnetic calorimeter (ECAL) energy resolution is critical for $\piz$ reconstruction, as a better energy resolution can significantly reduce the combinatorial background. The ECAL energy resolution is parameterized as
\begin{equation}
\frac{\sigma_E}{E}=\frac{\alpha}{\sqrt{E/\mathrm{GeV}}}\oplus\beta,
\label{eq:resolution}
\end{equation}
where $E$ is the photon energy, $\sigma_E$ is its resolution, $\alpha$ is the stochastic coefficient, and $\beta$ is the constant term. The symbol $\oplus$ denotes addition in quadrature. The reference setting in this study is $\alpha=0.03$ and $\beta=0.01$, corresponding to the nominal ECAL energy resolution implemented in the adopted CEPC detector card and consistent with the photon energy resolution requirement in the CEPC Reference Detector TDR~\cite{CEPC_RefTDR}. To investigate the dependence of $\piz$ reconstruction performance on the ECAL energy resolution, we vary the stochastic coefficient $\alpha$ from 0.01 to 0.25. For reference, the stochastic and constant terms are $2.8\%$ and $0.3\%$ for the CMS crystal ECAL~\cite{CMS:ECAL2013}, $9\%$--$10\%$ (at low $|\eta|$) and $0.7\%$ for the ATLAS ECAL design~\cite{ATLAS:EGamma2014}, and $16.5\%$ and $1.1\%$ for the CALICE Si--W sampling ECAL prototype~\cite{CALICE:SiW2009}.

\begin{figure}[tbp]
\centering
\includegraphics[width=\textwidth]{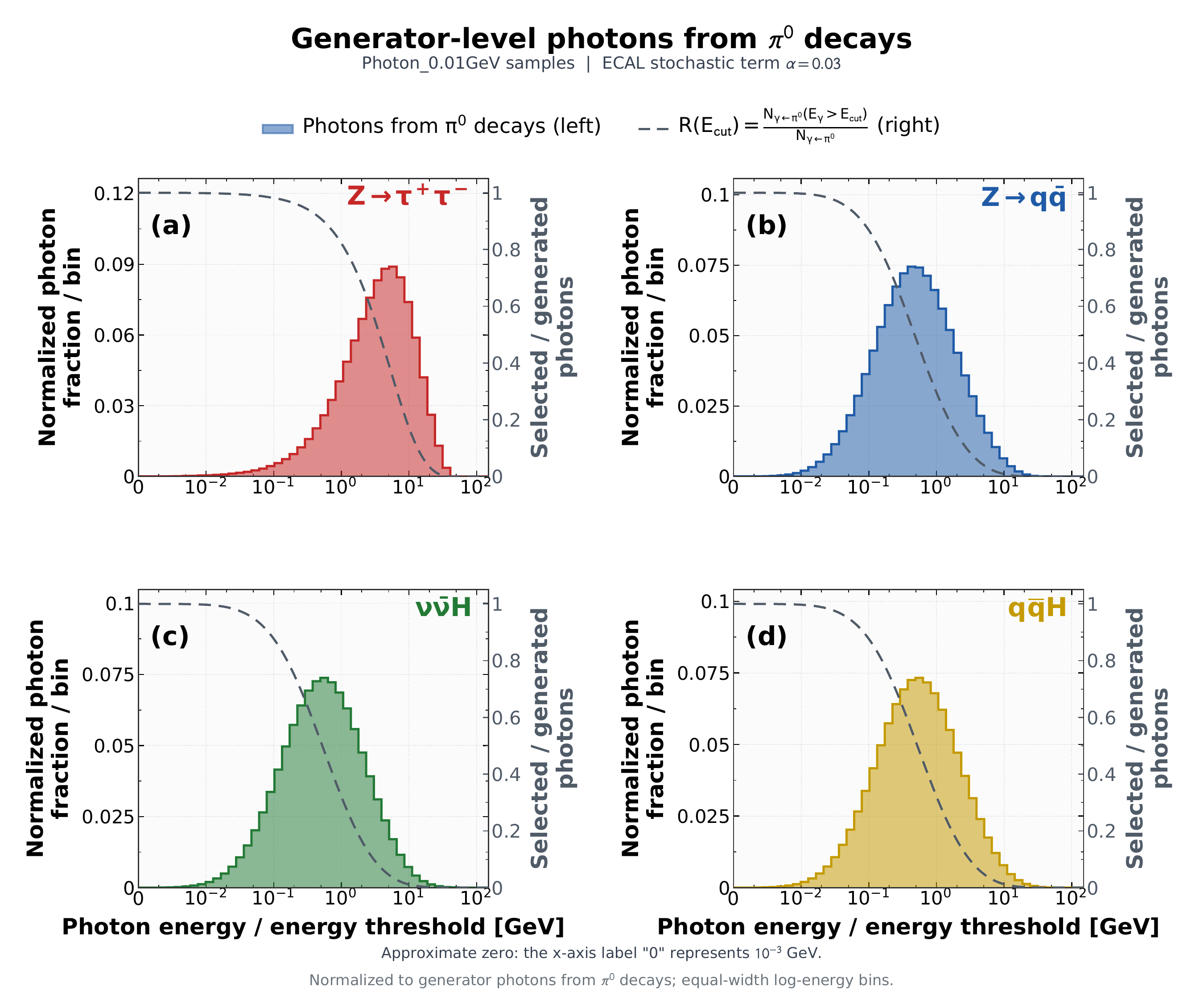}
\caption{Generator-level $\piz$ photon-energy distributions in (a) $Z\to\tau^+\tau^-$, (b) $Z\to q\bar q$, (c) $\nnh$ and (d) $\qqh$ at the reference ECAL setting ($\alpha=0.03$, $\beta=0.01$). The histograms show the normalized photon-energy distributions, while the dashed curves indicate the fraction of generated photons with $E_\gamma>E_{\rm cut}$ as a function of the energy threshold $E_{\rm cut}$.}
\label{fig:photon_energy}
\end{figure}

Another key parameter is the photon-energy threshold. In this study, it is set to $10~\mathrm{MeV}$ in Delphes to retain approximately $99\%$ of the generator-level photons from $\piz$ decays in all four processes, as shown in figure~\ref{fig:photon_energy}. The actual photon-energy threshold depends on the collision environment, the amount of upstream material, and the ECAL technology. The threshold of $10~\mathrm{MeV}$ is relatively optimistic compared with the selection thresholds used in $\piz$ and hadronic $\tau$ reconstruction at the LHC. For example, ATLAS uses a transverse-energy threshold of $100~\mathrm{MeV}$ to identify local energy maxima in the first ECAL layer~\cite{ATLAS:tauPFlow}, while CMS uses a transverse-momentum threshold of $500~\mathrm{MeV}$ for photons and electrons in its strip reconstruction algorithm~\cite{CMS:tauReco2016}. It is, however, closer to the minimum photon-energy threshold of $25~\mathrm{MeV}$ used in the BESIII barrel~\cite{BESIII:chic4pi0}.

\subsection{Candidate selection and performance measures}
\label{sec:metric}

By comparing reconstructed photons with Monte Carlo truth information, we identify photon pairs matched to the two distinct daughters of the same generated $\piz$ as signal, and all other pairs as background. We denote the number of correctly matched photon pairs before mass preselection by $N_{\rm true\ pair}^{\rm all}$, the number of correctly matched pairs after selection by $N_{\rm true\ pair}$, and the total number of selected pairs by $N_{\rm sel\ pair}$. The selection efficiency, purity, and their product are defined as
\begin{equation}
\varepsilon=\frac{N_{\rm true\ pair}}{N_{\rm true\ pair}^{\rm all}},
\qquad
P=\frac{N_{\rm true\ pair}}{N_{\rm sel\ pair}},
\qquad
\metric=\varepsilon P.
\label{eq:metric}
\end{equation}
The efficiency is defined relative to generated $\piz$ decays for which both daughter photons are reconstructed and truth-matched. No event-level exclusivity is imposed, so a reconstructed photon may appear in multiple selected pairs.

For each method, the selection criteria are optimized by maximizing $\metric$ on validation data, and the resulting working points are applied to independent test samples. The relative improvement of method $A$ over the classical baseline (MW) is defined as
\begin{equation}
G_A=\frac{\metric_A-\metric_{\rm MW}}{\metric_{\rm MW}}.
\label{eq:gain}
\end{equation}

\begin{figure}[tbp]
    \centering
    \includegraphics[width=\textwidth]{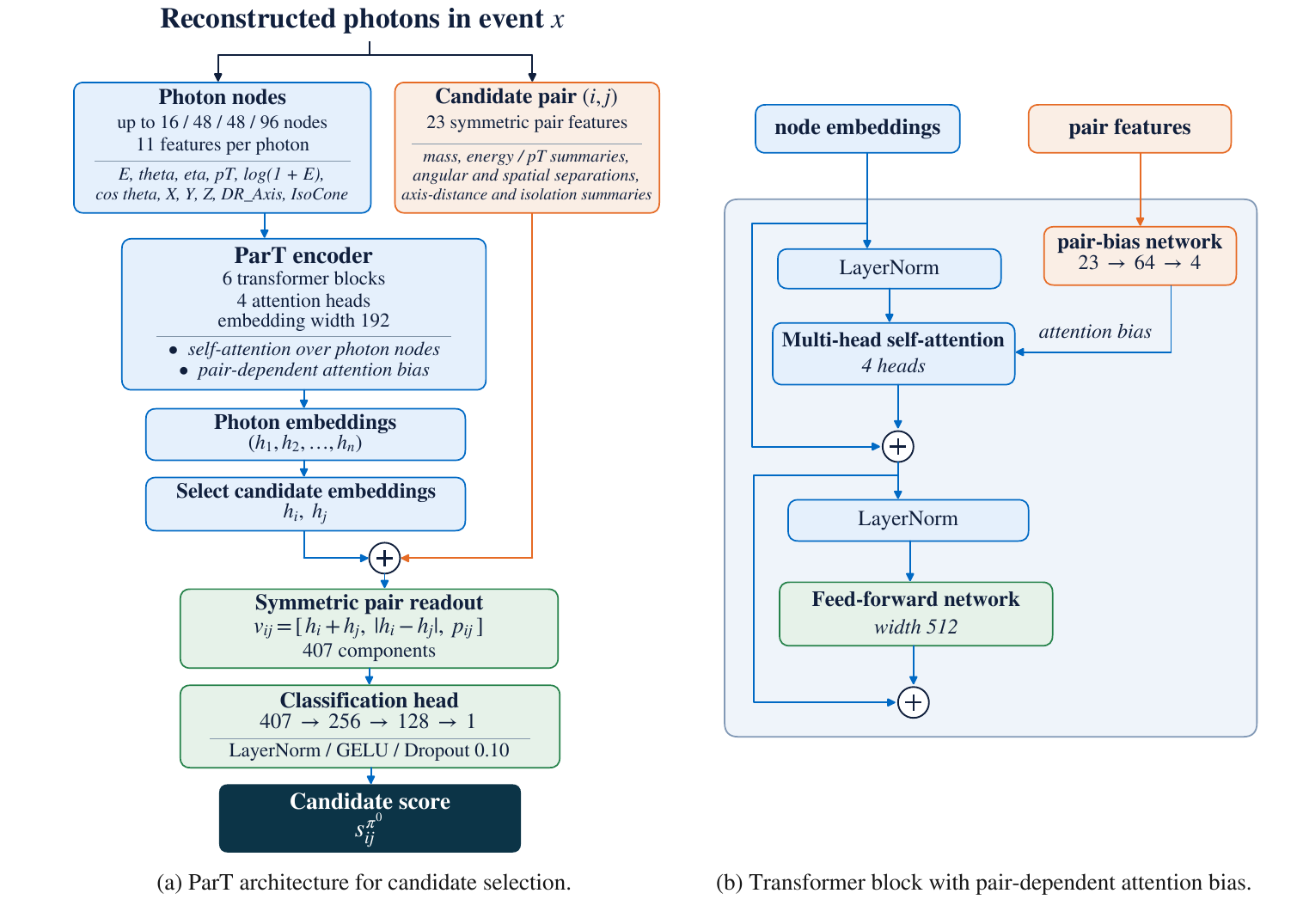}
    \caption{Architecture of the Particle Transformer (ParT) for $\piz$ candidate selection: (a) the overall architecture with symmetric candidate-pair readout and (b) the transformer block with pair-dependent attention bias.}
    \label{fig:workflow}
\end{figure}

For each candidate photon pair, the classical baseline (MW) uses only its invariant mass, while BDT and MLP use 21 and 23 pair-level features, respectively, including the diphoton invariant mass, photon energies, transverse momenta, angular separations, and variables related to photon isolation and the event axis. Through the latter variables, BDT and MLP implicitly incorporate information from the rest of the event. PN and ParT further incorporate correlations among reconstructed photons through local neighborhood interactions and self-attention, respectively. The encoded representations of the two photons forming the candidate are then combined with the pair-level features to obtain a selection score. Figure~\ref{fig:workflow} illustrates the architecture of ParT. The input features and further implementation details are provided in appendix~\ref{sec:implementation}.

All four machine-learning methods start with a loose diphoton mass preselection applied to the candidate photon pairs to ensure a high signal efficiency before classification. For each process, all four machine-learning methods use the same preselection, with the same midpoint as the optimized mass interval of the classical baseline (MW) and a half-width enlarged by factors of 5, 6, 6, and 10 for $Z\to\tau^+\tau^-$, $Z\to q\bar q$, $\nnh$, and $\qqh$, respectively. At the reference ECAL setting, these loose mass windows retain approximately $99\%$ of the reconstructable signal pairs while rejecting part of the combinatorial background. The machine-learning methods then further discriminate between signal and combinatorial background. Further technical details are provided in appendix~\ref{sec:technical_details}.

\section{Results}
\label{sec:results}

\subsection{Candidate-selection performance}
\label{sec:reference_performance}
\label{sec:event_displays}

We first compare the performance of the five methods at the reference ECAL setting ($\alpha=0.03$, $\beta=0.01$). For each process, the machine-learning models are trained on $10^6$ events. The selection thresholds and mass-window boundaries are optimized using 20,000 validation events, and the performance is evaluated on 30,000 independent test events. Table~\ref{tab:bootstrap_results} summarizes the efficiency--purity product $\metric$ for the five methods in the four processes. The corresponding efficiencies, purities, and statistical uncertainties are reported in tables~\ref{tab:detail_effpurity} and~\ref{tab:breakdown_results}.

\begin{table}[t]
\centering
\caption{Efficiency--purity product $\metric$ (\%) at the reference ECAL
setting ($\alpha=0.03$, $\beta=0.01$). Bootstrap uncertainties are given
in Table~\ref{tab:breakdown_results}.}
\label{tab:bootstrap_results}
\small
\setlength{\tabcolsep}{8pt}
\renewcommand{\arraystretch}{1.14}
\begin{tabular}{@{}lcccc@{}}
\toprule
Method & $Z\to\tau^+\tau^-$ & $Z\to q\bar q$ & $\nnh$ & $\qqh$ \\
\midrule
MW   & 53.93 & 32.94 & 27.47 & 23.05 \\
BDT  & 69.46 & 41.40 & 35.64 & 33.18 \\
MLP  & 69.48 & 41.61 & 35.80 & 33.52 \\
PN   & 82.24 & 56.62 & 48.04 & 44.89 \\
ParT & 86.32 & 66.21 & 56.76 & 52.57 \\
\midrule
ParT gain over MW (\%) & 60.06 & 101.00 & 106.61 & 128.05 \\
\bottomrule
\end{tabular}
\end{table}

All four machine-learning methods outperform the classical baseline (MW). BDT and MLP achieve similar performance, with differences in $\metric$ of less than 0.5 percentage in each process and relative gains of 26\%--45\% over the classical baseline. PN and ParT achieve substantially larger relative gains of 53\%--95\% and 60\%--128\%, respectively, across the four processes. These gains are approximately 1.8--3.9 times those achieved by BDT and MLP in the corresponding processes. Among the four machine-learning methods, ParT achieves the highest $\metric$ in all four processes. The gains obtained with PN and ParT arise mainly from higher purity rather than increased selection efficiency. For example, in $\nnh$, purity increases from 35.37\% for the classical baseline to 67.66\% for PN and 76.51\% for ParT, while the corresponding efficiencies are 77.66\%, 71.00\%, and 74.18\%. For PN and ParT, the absolute increases in $\metric$ over the classical baseline range from 20.57 to 33.27 percentage. The individual one-standard-deviation bootstrap uncertainties on $\metric$ in table~\ref{tab:breakdown_results} range from 0.05 to 0.33 percentage.

\begin{figure}[!tbp]
\centering
\eventdisplay{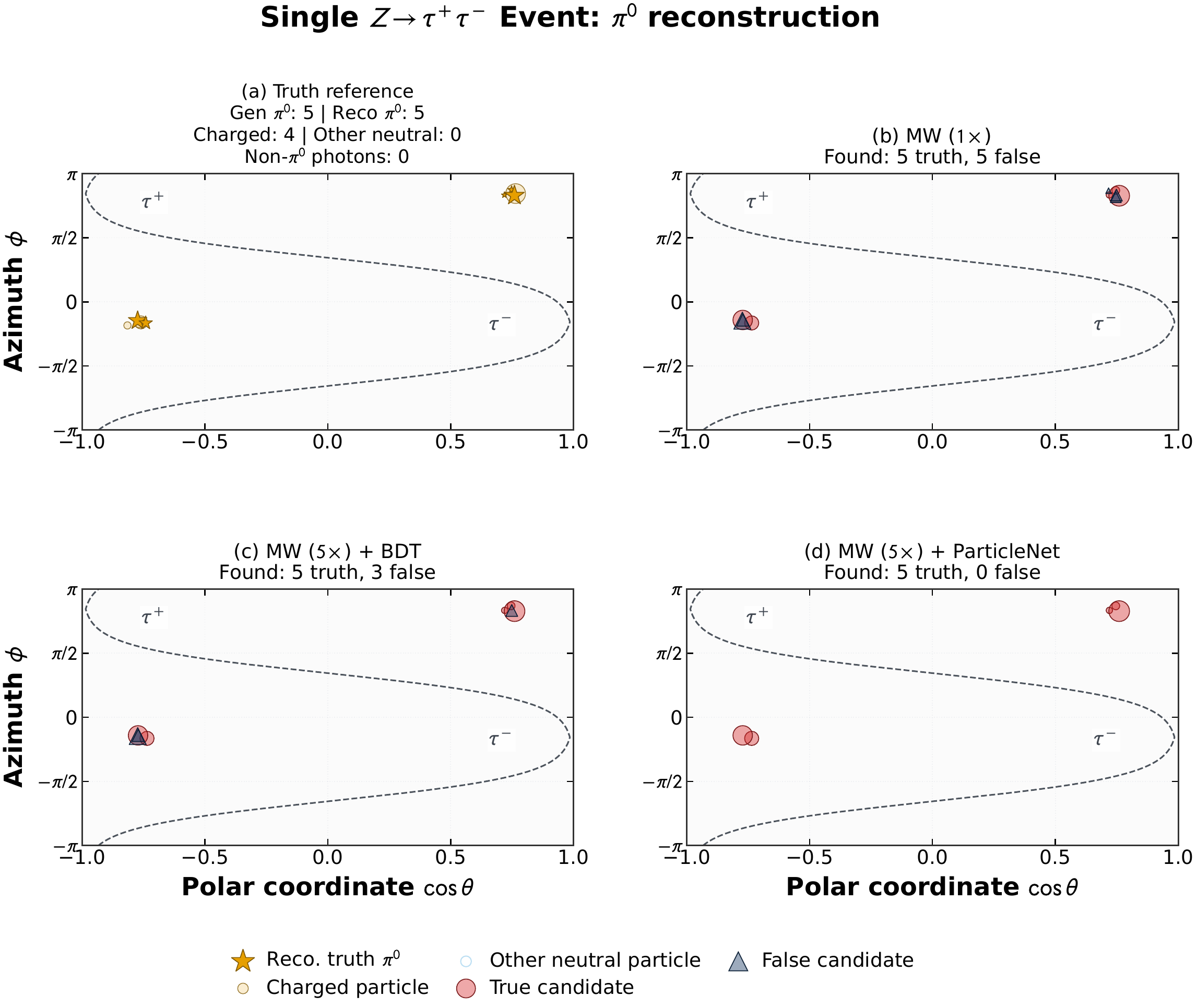}
\caption{An illustrative $Z\to\tau^+\tau^-$ event at the reference ECAL setting ($\alpha=0.03$, $\beta=0.01$). Panel (a) shows the generator-level reference; panels (b), (c), and (d) show candidates selected by MW, BDT, and PN, respectively. Orange stars denote generated $\piz$ mesons with reconstructable daughter-photon pairs, red circles correctly matched candidates, and blue triangles incorrectly matched candidates. Translucent orange circles denote charged particles, and open light-blue circles other neutral particles. The labels $1\times$ and $5\times$ indicate the optimized mass window of the classical baseline and the loose mass preselection used before BDT and PN classification, respectively; the multiplier applies to the window half-width at fixed center. A minimum marker size is used for visibility. All generated $\piz$ mesons in this event have reconstructable daughter-photon pairs. MW, BDT, and PN select 5, 5, and 5 correctly matched pairs, and 5, 3, and 0 incorrect pairs, respectively.}
\label{fig:pn_event_display_Ztautau}
\end{figure}

\begin{figure}[!tbp]
\centering
\eventdisplay{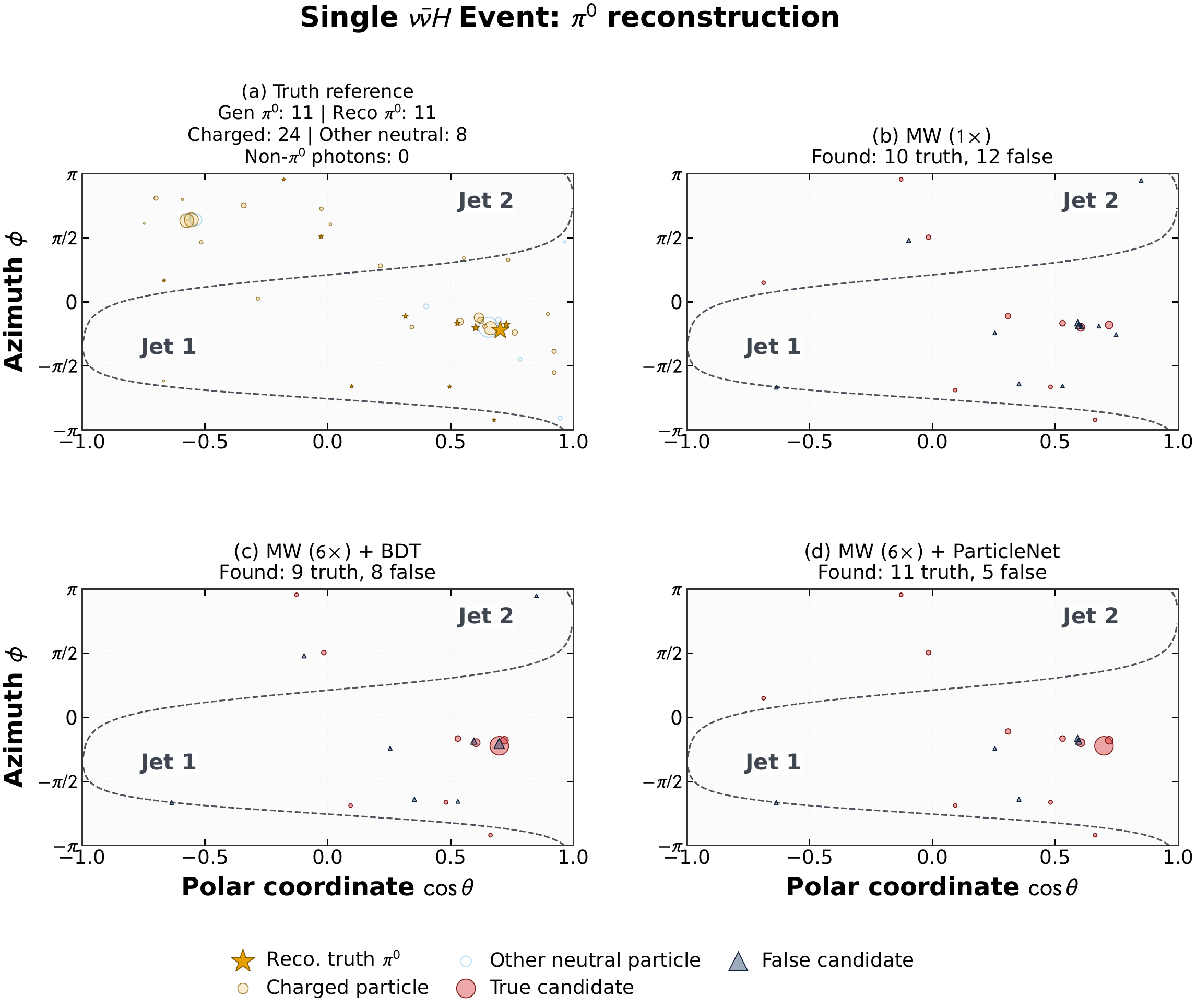}
\caption{An illustrative $\nnh$ event at the reference ECAL setting ($\alpha=0.03$, $\beta=0.01$). The panel arrangement and marker definitions follow those in figure~\ref{fig:pn_event_display_Ztautau}, with a $6\times$ mass preselection for BDT and PN. All generated $\piz$ mesons in this event have reconstructable daughter-photon pairs. MW, BDT, and PN select 10, 9, and 11 correctly matched pairs, and 12, 8, and 5 incorrect pairs, respectively.}
\label{fig:pn_event_display_vvH}
\end{figure}

Figures~\ref{fig:pn_event_display_Ztautau} and~\ref{fig:pn_event_display_vvH} show two illustrative events from the $Z\to\tau^+\tau^-$ and $\nnh$ samples. Directions are shown in $(\cos\theta,\phi)$, with each candidate placed along the combined momentum of its two photons. The generator-level panels show the $\piz$ mesons with reconstructable daughter-photon pairs; all generated $\piz$ mesons in these two examples satisfy this condition. The remaining panels show correctly and incorrectly matched photon pairs selected by MW, BDT, and PN. In both examples, PN selects fewer incorrect pairs than the classical baseline and BDT while maintaining a comparable or higher number of correctly matched pairs. These examples are consistent with the improved purity reported in table~\ref{tab:detail_effpurity} and the resulting gain in $\metric$ shown in table~\ref{tab:bootstrap_results}. Additional event displays for $Z\to q\bar q$ and $\qqh$ are provided in appendix~\ref{sec:additional_event_displays}.


\subsection{Dependence on ECAL resolution}
\label{sec:resolution_results}

\begin{figure}[t]
\centering
\includegraphics[width=\textwidth]{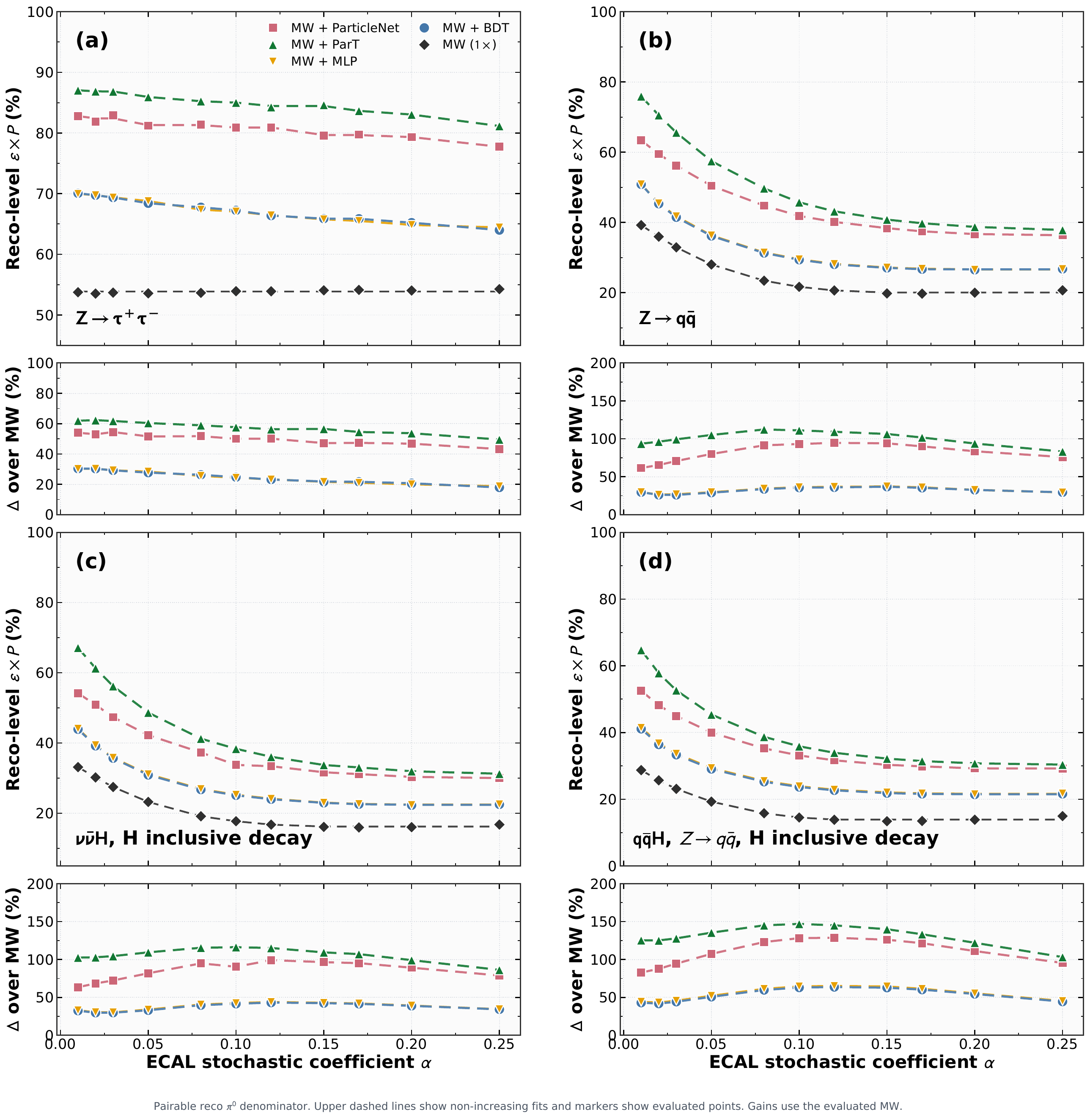}
\caption{Validation-sample efficiency--purity product $\metric$ (upper panels) and relative gain over the classical baseline (lower panels), using $9.5\times10^5$ training events, as functions of the ECAL stochastic coefficient $\alpha$, with $\beta=0.01$: (a) $Z\to\tau^+\tau^-$, (b) $Z\to q\bar q$, (c) $\nnh$, and (d) $\qqh$. Curves connect the tested settings.}
\label{fig:resolution}
\end{figure}

Figure~\ref{fig:resolution} compares the performance of the five methods as a function of the ECAL stochastic coefficient $\alpha$. All four machine-learning methods maintain their performance advantage over the classical baseline (MW) throughout the investigated range of $\alpha$, with relative gains of approximately 20\%--65\%, 20\%--65\%, 45\%--130\%, and 50\%--150\% for BDT, MLP, PN, and ParT, respectively, across the four benchmark processes. For each process, the relative gains vary smoothly with $\alpha$, and the performance hierarchy among the methods is maintained. For $Z\to\tau^+\tau^-$, $\metric$ shows only a weak dependence on $\alpha$, whereas a more pronounced decrease is observed in the other three processes as $\alpha$ increases. The weaker dependence in $Z\to\tau^+\tau^-$ is consistent with its much lower $\piz$ multiplicity and fewer competing photon-pair assignments (figure~\ref{fig:multiplicity}). The mass-window boundaries of the classical baseline and the selection thresholds of the machine-learning methods are reoptimized, and the classifiers are retrained at each resolution setting.

In $Z\to\tau^+\tau^-$, the relative gains over the classical baseline (MW) gradually decrease with increasing $\alpha$, whereas in the other three processes, they generally increase initially and then decrease, reaching or approaching their maximum values at $\alpha\approx0.05$--$0.12$. Even at $\alpha=0.25$, ParT still improves $\metric$ relative to the classical baseline by $83.5\%$, $86.6\%$, and $103.2\%$ in $Z\to q\bar q$, $\nnh$, and $\qqh$, respectively.

\subsection{Scaling: performance dependence on training data size}
\label{sec:scaling}

Scaling laws in machine learning, relating model performance to training-sample size, have been observed in a wide range of deep-learning tasks~\cite{Hestness:2017scaling,Rosenfeld:2020scaling}. Models with different architectures and input representations may exhibit different scaling behaviors and require different amounts of training data to approach performance saturation~\cite{Viering:2023learning}. We therefore investigate the learning curves of the four machine-learning methods by evaluating $\metric$ as a function of the number of training events, keeping the ECAL energy resolution fixed at the reference setting ($\alpha=0.03$, $\beta=0.01$).

\begin{figure}[t]
\centering
\includegraphics[width=\textwidth]{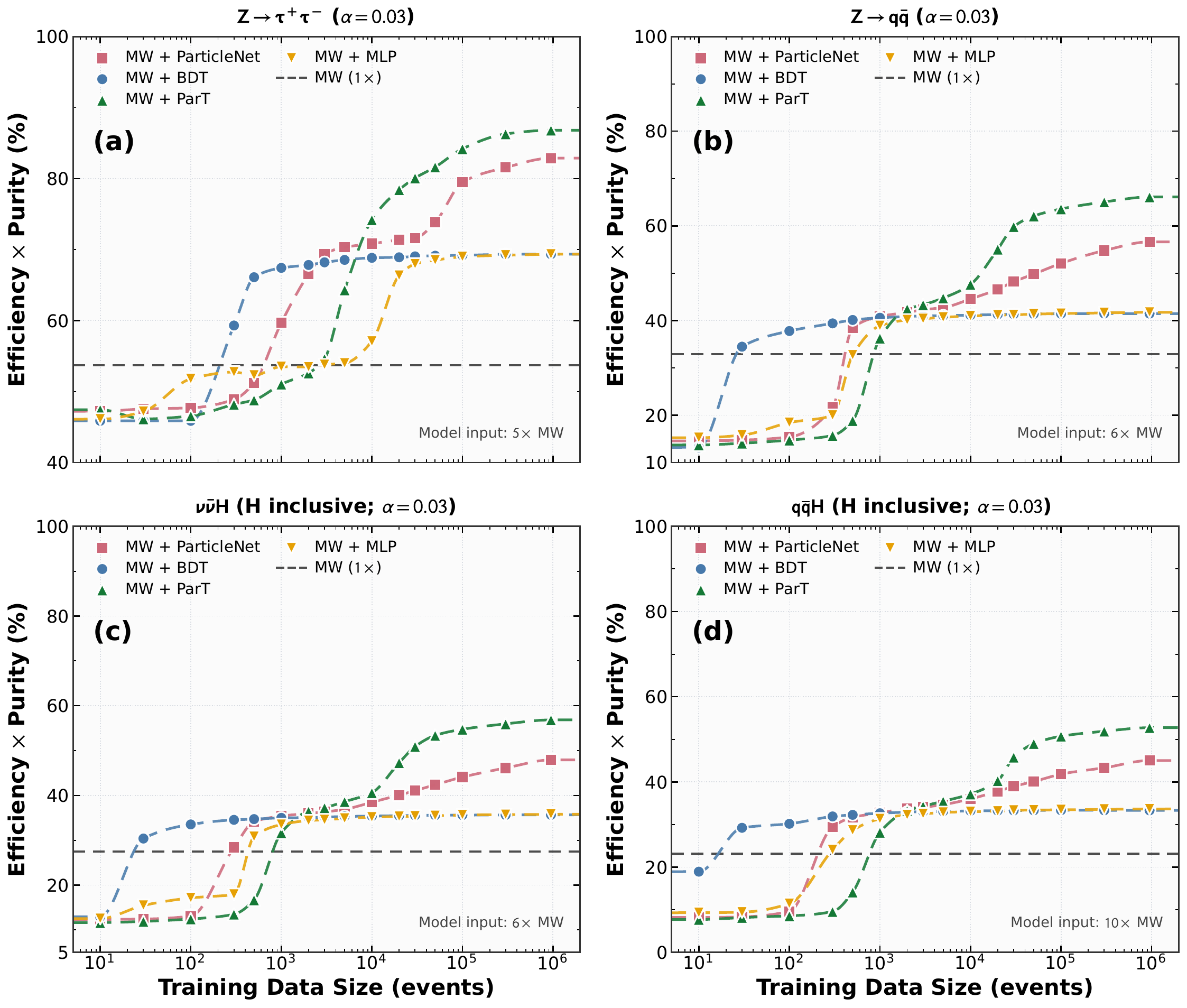}
\caption{Validation-sample efficiency--purity product $\metric$ as a function of the number of training events at $\alpha=0.03$ and $\beta=0.01$: (a) $Z\to\tau^+\tau^-$, (b) $Z\to q\bar q$, (c) $\nnh$, and (d) $\qqh$. Dashed horizontal lines indicate the classical baseline (MW) for each process. PN and ParT are trained for 50 epochs, while MLP uses up to 40 epochs with early stopping. Curves are obtained using shape-preserving cubic Hermite interpolation in $x=\log_{10}(N_{\mathrm{train}})$ and serve as guides to the eye.}
\label{fig:scaling}
\end{figure}

Figure~\ref{fig:scaling} shows the learning curves of the four machine-learning methods for the four benchmark processes. A general S-shaped behavior is observed as a function of the logarithm of the training-sample size. With very limited training statistics, the models have little discriminating power between signal and combinatorial background, and their performance can be worse than that of the classical baseline (MW). At the smallest training-sample size considered ($10$ events), the performance is close to the trivial performance obtained with the loose mass preselection alone. As the training-sample size increases, the learning curves exhibit phase-transition-like behavior, entering a regime of rapid performance improvement, followed by a regime of slower growth or saturation. The characteristic training-sample sizes associated with these transitions depend on the model and physics process. BDT approaches its plateau with relatively few training events, while MLP requires larger samples to reach a similar performance. PN and ParT continue to improve over a much wider range of training-sample sizes. For example, in $Z\to q\bar q$, $\metric$ for ParT increases from $47.6\%$ at $10^4$ training events to $66.1\%$ at $\sim10^6$ events, whereas BDT and MLP reach approximately $41\%$--$42\%$. These results demonstrate that PN and ParT benefit more from increased training statistics within the investigated range.

For the largest training sample, ParT achieves the highest $\metric$ in all four processes. Its performance in $\qqh$ reaches approximately $53\%$, below the $66\%$ and $57\%$ obtained in $Z\to q\bar q$ and $\nnh$, respectively. The $\qqh$ sample contains the associated $Z\to q\bar q$ decay together with the inclusive Higgs decay. When the Higgs boson decays to $b\bar b$, $c\bar c$, or $gg$, the two decay systems can give four-jet final states~\cite{Ruan:2018yrh}, introducing additional photon-pair combinations between them. The lower performance in $\qqh$ is consistent with its greater combinatorial complexity and higher $\piz$ multiplicity (figure~\ref{fig:multiplicity}).

Beyond the general S-shaped behavior, two-stage S-shaped learning curves are observed in several cases for PN and ParT, as shown in figure~\ref{fig:scaling}. This double-S structure is particularly pronounced for PN in $Z\to\tau^+\tau^-$ and ParT in $Z\to q\bar q$ and $\nnh$, where two stages of rapid improvement in $\metric$ are separated by a region of slower growth. Similar structures are also visible in other learning curves, although the prominence of the intermediate region and the training-sample sizes at which the two stages occur vary among models and processes.

\begin{figure}[!htbp]
\centering
\begin{minipage}[t]{0.49\textwidth}
    \centering
    \includegraphics[width=\linewidth]{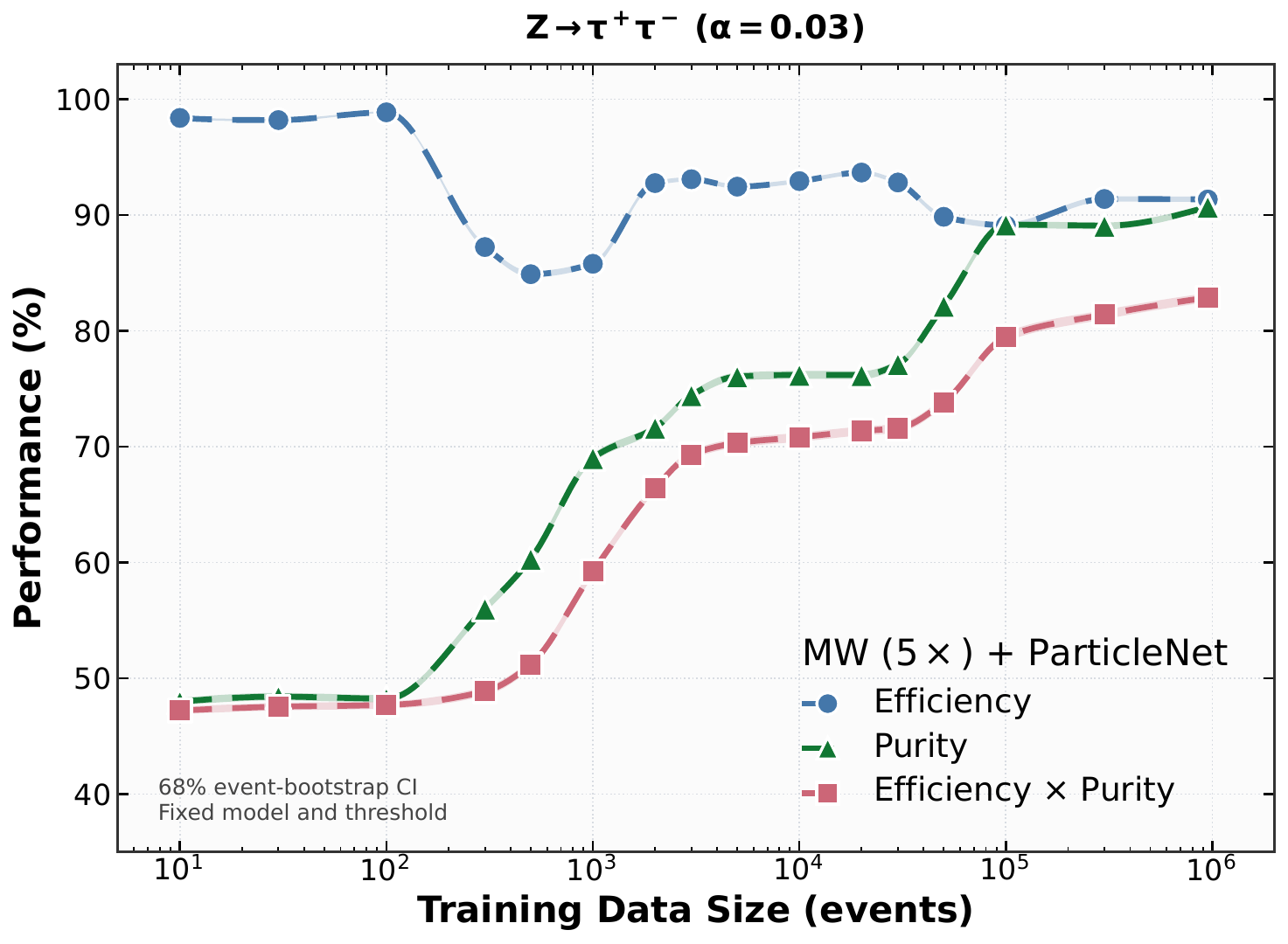}
\end{minipage}
\hfill
\begin{minipage}[t]{0.49\textwidth}
    \centering
    \includegraphics[width=\linewidth]{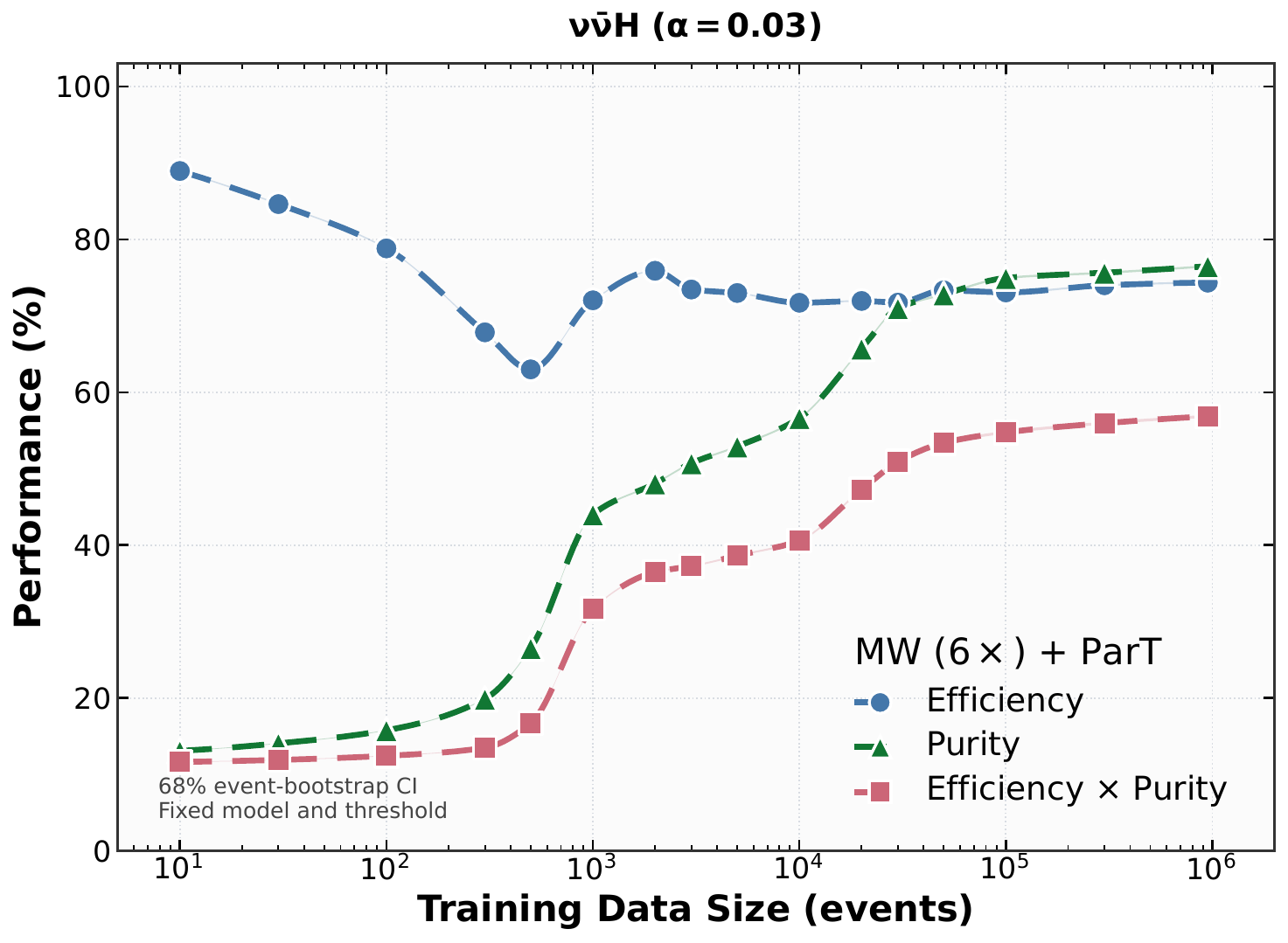}
\end{minipage}
\caption{Efficiency, purity and their product as functions of training-sample size at $\alpha=0.03$ and $\beta=0.01$: (a) PN in $Z\to\tau^+\tau^-$ and (b) ParT in $\nnh$. The results include pointwise 68\% event-level bootstrap intervals on validation samples, conditional on fixed models and score thresholds.}
\label{fig:double_s_diagnostic}
\end{figure}

To further characterize the double-S structure, figure~\ref{fig:double_s_diagnostic} shows the selection efficiency and purity as functions of the training-sample size for PN in $Z\to\tau^+\tau^-$ and ParT in $\nnh$. For PN in $Z\to\tau^+\tau^-$, purity increases from approximately 48\% to 76\% during the first stage and reaches about 90\% after the second stage, without a corresponding increase in efficiency. A similar behavior is observed for ParT in $\nnh$, where purity increases from approximately 13\% to 50\%--56\% during the first stage and subsequently reaches about 76\%. In both cases, the two-stage increase in $\metric$ is primarily driven by improved rejection of combinatorial background rather than increased selection efficiency. Since the checkpoint and score threshold are chosen to maximize $\metric=\varepsilon P$ at each training size, an increase in purity can compensate for a decrease in efficiency at the selected working point.

\FloatBarrier

\FloatBarrier
\section{Summary}
\label{sec:summary}

In this work, we have developed four machine-learning methods for pairing reconstructed photons into $\piz$ candidates and evaluated their performance in benchmark physics processes at future $Z$ and Higgs factories. At the reference ECAL energy resolution, all four methods outperform the classical baseline (MW), with PN and ParT achieving relative improvements in the efficiency--purity product $\metric$ of $53\%$--$95\%$ and $60\%$--$128\%$, respectively, primarily through improved rejection of combinatorial background. The performance advantage is maintained across the investigated range of ECAL stochastic coefficients, covering those typical of homogeneous crystal and sampling calorimeters.

The learning curves exhibit a general S-shaped behavior as a function of the logarithm of the training-sample size, progressing from nearly trivial performance with $10$ training events through rapid improvement towards slower growth or saturation. BDT approaches saturation with relatively few training events, while MLP requires larger samples to reach a similar performance. PN and ParT continue to improve with increasing training-sample size, and ParT achieves the best performance in all four processes at the largest training-sample size. Two-stage S-shaped learning curves are observed in several cases for PN and ParT. In both cases examined in detail, the two stages of improvement are primarily driven by increases in purity.

The classical baseline (MW) provides a simple and widely used reference for $\piz$ reconstruction. More elaborate methods have been developed to reduce combinatorial background, including candidate ranking based on mass-constrained fits and photon opening angles~\cite{Beddall:2002pi0}, and photon-energy-based ranking with exclusive photon assignment~\cite{Belle:2018tauPolarization}. The former method improves the efficiency--purity product by approximately $8\%$ for simulated ALEPH events, while the signal and background retention rates reported by Belle imply an upper bound of approximately $26\%$ on the relative improvement from its ranking procedure. The relative gains of $53$--$95\%$ and $60$--$128\%$ achieved by PN and ParT, respectively, in this study are substantially larger than these values.

Future studies will explore alternative model architectures and incorporate information from charged particles and global event properties to develop more direct methods for particle reconstruction and decay-mode identification. Starting from $\piz$ reconstruction, these methods can be extended to $\rho^\pm$, $\tau$, $K_S^0$, $\Lambda$, $\phi$, $K^*$, and eventually to heavy-flavor hadrons such as $D$ and $B$ mesons with more complex decay chains. The long-term goal is to reconstruct unstable particles produced in high-energy collisions, particularly those emerging from hadronization, and recover their decay histories. This bottom-up reconstruction strategy, referred to as ``from leaves to the tree''~\cite{Ruan:2026Marseille,Kahn:2022tree}, views reconstructed final-state particles as the leaves of a tree, with their parent particles and decay chains forming its branches and trunk. Such a strategy could improve the precision and discovery potential of flavor physics at future $Z$ and Higgs factories and provide new observables for studies of hadronization and non-perturbative QCD.

\section*{Acknowledgments}
This work is supported by the National Key R\&D Program of China under Grant No.\ 2024YFA1610603 and the NSFC-funded International Collaboration Fund for Research Teams under Grant No.~W2441004. This work is partially supported by the China Postdoctoral Science Foundation under Grant No.~2025M773403.

\appendix
\section{Additional reference-setting results}
\label{sec:detailed_results}

The efficiencies and purities are given in table~\ref{tab:detail_effpurity}; products and relative gains are given in table~\ref{tab:breakdown_results}. The quoted uncertainties on $\metric$ reflect finite test-sample statistics and do not include run-to-run training variations.

\begin{table}[!htbp]
\centering
\caption{Efficiency $\varepsilon$ and purity $P$ at the reference ECAL setting. Values are fractions, not percentages; the multipliers in parentheses indicate the enlargement of the mass-window half-width relative to the classical baseline for the machine-learning methods.}
\label{tab:detail_effpurity}
\small
\setlength{\tabcolsep}{7pt}
\renewcommand{\arraystretch}{1.12}
\begin{tabular}{@{}lccccc@{}}
\toprule
\multicolumn{6}{c}{\textit{Efficiency $\varepsilon$}} \\
\midrule
Event sample & MW & BDT & MLP & PN & ParT \\
\midrule
$Z\to\tau^+\tau^-$ ($5\times$) & 0.8461 & 0.9460 & 0.9520 & 0.9124 & 0.9099 \\
$Z\to q\bar q$ ($6\times$) & 0.8067 & 0.8284 & 0.8188 & 0.7650 & 0.7984 \\
$\nnh$ ($6\times$) & 0.7766 & 0.8012 & 0.7962 & 0.7100 & 0.7418 \\
$\qqh$ ($10\times$) & 0.7738 & 0.7378 & 0.7512 & 0.6885 & 0.7045 \\
\midrule
\multicolumn{6}{c}{\textit{Purity $P$}} \\
\midrule
Event sample & MW & BDT & MLP & PN & ParT \\
\midrule
$Z\to\tau^+\tau^-$ ($5\times$) & 0.6374 & 0.7343 & 0.7298 & 0.9014 & 0.9487 \\
$Z\to q\bar q$ ($6\times$) & 0.4084 & 0.4998 & 0.5082 & 0.7401 & 0.8293 \\
$\nnh$ ($6\times$) & 0.3537 & 0.4448 & 0.4496 & 0.6766 & 0.7651 \\
$\qqh$ ($10\times$) & 0.2979 & 0.4497 & 0.4462 & 0.6521 & 0.7462 \\

\bottomrule
\end{tabular}
\end{table}

\begin{table}[!htbp]
\centering
\caption{Efficiency--purity product $\metric$ and relative gain (\%) over the classical baseline (MW) at $\alpha=0.03$ and $\beta=0.01$. A value such as $0.5393(30)$ denotes $0.5393\pm0.0030$; the parentheses give one standard deviation from 20,000 event-level bootstrap replicas of 30,000 independent test events per process. Gains are quoted without paired-bootstrap uncertainties.}
\label{tab:breakdown_results}
\small
\setlength{\tabcolsep}{7pt}
\renewcommand{\arraystretch}{1.13}
\begin{tabular}{@{}lccccc@{}}
\toprule
\multicolumn{6}{c}{\textit{Efficiency--purity product $\metric$}} \\
\midrule
Event sample & MW & BDT & MLP & PN & ParT \\
\midrule
$Z\to\tau^+\tau^-$ & $0.5393(30)$ & $0.6946(30)$ & $0.6948(29)$ & $0.8224(33)$ & $0.8632(32)$ \\
$Z\to q\bar q$ & $0.3294(9)$ & $0.4140(10)$ & $0.4161(10)$ & $0.5662(14)$ & $0.6621(16)$ \\
$\nnh$ & $0.2747(7)$ & $0.3564(9)$ & $0.3580(9)$ & $0.4804(13)$ & $0.5676(15)$ \\
$\qqh$ & $0.2305(5)$ & $0.3318(7)$ & $0.3352(7)$ & $0.4489(10)$ & $0.5257(12)$ \\
\midrule
\multicolumn{6}{c}{\textit{Gain (\%)}} \\
\midrule
Event sample & MW & BDT & MLP & PN & ParT \\
\midrule
$Z\to\tau^+\tau^-$ & -- & $28.81$ & $28.84$ & $52.50$ & $60.06$ \\
$Z\to q\bar q$ & -- & $25.68$ & $26.32$ & $71.88$ & $101.00$ \\
$\nnh$ & -- & $29.72$ & $30.31$ & $74.85$ & $106.61$ \\
$\qqh$ & -- & $43.95$ & $45.41$ & $94.74$ & $128.05$ \\

\bottomrule
\end{tabular}
\end{table}
\FloatBarrier
\section{Implementation and evaluation}
\label{sec:technical_details}

\subsection{Photon-pair construction and truth matching}
\label{sec:truth_association}

Reconstructed photons are combined into unordered candidate pairs $(i,j)$, with invariant mass
\begin{equation}
m_{ij}^2=2E_iE_j\left(1-\cos\theta_{ij}\right),
\label{eq:mass}
\end{equation}
where $\theta_{ij}$ is the opening angle between the two photons. In the simulated samples, each reconstructed photon retains the index of its associated generator-level photon, and each generator-level photon corresponds to at most one reconstructed photon. A candidate is labeled as signal if its two photons are matched to distinct generator-level photons originating from the same $\piz\to\gamma\gamma$ decay. All other pairs are labeled as background. The generator-level indices and parent information are used only for truth labeling and performance evaluation, and are not provided as classifier inputs. No event-level exclusivity requirement is imposed, so a reconstructed photon may appear in multiple selected pairs.

Because the photon--MC association is one-to-one, each generated $\piz\to\gamma\gamma$ decay contributes one true pair if both daughter photons are reconstructed and matched, and none otherwise. Thus $N_{\rm true\ pair}^{\rm all}$ counts these available pairs before mass preselection, not all generated $\piz$ mesons. The efficiency in section~\ref{sec:metric} measures their retention after candidate selection.

\subsection{Mass-window selection and preselection}
\label{sec:mass_preselection}

The ECAL-resolution scan fixes $\beta=0.01$ and uses
\begin{equation*}
\begin{split}
\alpha\in\{&0.01,0.02,0.03,0.05,0.08,0.10,\\
       &0.12,0.15,0.17,0.20,0.25\}.
\end{split}
\end{equation*}
All other detector-card parameters are held fixed.

\noindent\textbf{Classical baseline (MW).}
Conventional resolved-$\piz$ reconstruction selects photon pairs with invariant masses close to the nominal $\piz$ mass. For example, Belle used $118<m_{\gamma\gamma}<150~\mathrm{MeV}$, followed by a mass-constrained fit, in its study of $B^0\to\omega\omega$~\cite{Belle:omegaomega}. The classical baseline (MW) uses an optimized diphoton mass window, selecting candidates that satisfy
\begin{equation}
m_{\mathrm{min}}<m_{\gamma\gamma}<m_{\mathrm{max}}.
\end{equation}
The window boundaries are optimized separately for each sample and ECAL resolution setting by maximizing $\metric$ on validation data. They are then fixed for evaluation on independent test events. This provides a mass-only reference without the additional constraints of a complete event-level reconstruction.

\noindent\textbf{Loose mass preselection.}
For the machine-learning methods, candidate pairs first pass a looser mass window before classification. This window is obtained by multiplying the half-width of the optimized mass window of the classical baseline by a factor $k$, while keeping its center unchanged. A wider window retains more signal pairs but also introduces additional combinatorial background and increases the computational cost. Based on validation-sample performance, we use $k=5$, $6$, $6$, and $10$ for $Z\to\tau^+\tau^-$, $Z\to q\bar q$, $\nnh$, and $\qqh$, respectively. At the reference ECAL setting ($\alpha=0.03$, $\beta=0.01$), these windows retain approximately $99\%$ of the reconstructable signal pairs before classification.

All four learned methods use the same loose window within each process. Its multiplier is fixed across the ECAL scan, while the mass-window boundaries of the classical baseline are reoptimized at each resolution. Parameters are selected on validation data; pairs excluded by preselection cannot be recovered.

\subsection{Input features and model architectures}
\label{sec:implementation}

The pair-level inputs are model-dependent. BDT and PN use the same 21 features, all symmetric under exchange of the two photons (table~\ref{tab:pair_features}). MLP uses 23 features that retain the individual quantities of the two photons without explicit symmetrization, while ParT uses a different set of 23 symmetric features (table~\ref{tab:model_pair_features}). Each distinct photon pair is constructed only once in all five methods. PN and ParT also encode reconstructed photons as nodes, using the 11 variables in table~\ref{tab:photon_features}. For ParT, these are derived from the nine stored quantities $(p_x,p_y,p_z,E,X,Y,Z,\mathrm{DR}_{\rm Axis},\mathrm{IsoCone})$.

\begin{table}[t]
\centering
\caption{The 21 symmetric candidate-pair features used by BDT and PN. Here $i$ and $j$ denote the two photons, $E_{ij}=E_i+E_j$, and $p_{T,ij}$ and $\eta_{ij}$ refer to their combined momentum. The \texttt{DR\_Axis} and \texttt{IsoCone} entries are reconstructed-photon variables.}
\label{tab:pair_features}
\footnotesize
\setlength{\tabcolsep}{5pt}
\renewcommand{\arraystretch}{1.07}
\begin{tabular}{@{}r p{0.21\textwidth} p{0.65\textwidth}@{}}
\toprule
No. & Feature & Definition \\
\midrule
1 & \texttt{pair\_mass} & Diphoton invariant mass $m_{ij}$ (\texttt{M\_pair}). \\
2 & \texttt{pair\_energy} & Diphoton energy $E_{ij}$ (\texttt{E\_pair}). \\
3 & \texttt{pair\_angle} & Opening angle $\theta_{ij}$ (\texttt{Angle\_pair}). \\
4 & \texttt{max\_E} & $\max(E_i,E_j)$. \\
5 & \texttt{min\_E} & $\min(E_i,E_j)$. \\
6 & \texttt{abs\_delta\_E} & $|E_i-E_j|$. \\
7 & \texttt{max\_E\_frac} & $\max(E_i,E_j)/E_{ij}$. \\
8 & \texttt{min\_E\_frac} & $\min(E_i,E_j)/E_{ij}$. \\
9 & \texttt{eta\_mean} & $(\eta_i+\eta_j)/2$. \\
10 & \texttt{abs\_delta\_eta} & $|\eta_i-\eta_j|$. \\
11 & \texttt{delta\_r} & $\Delta R_{ij}=\sqrt{(\eta_i-\eta_j)^2+(\Delta\phi_{ij})^2}$. \\
12 & \texttt{max\_pt} & $\max(p_{T,i},p_{T,j})$. \\
13 & \texttt{min\_pt} & $\min(p_{T,i},p_{T,j})$. \\
14 & \texttt{max\_pt\_frac} & $\max(p_{T,i},p_{T,j})/(p_{T,i}+p_{T,j})$. \\
15 & \texttt{min\_pt\_frac} & $\min(p_{T,i},p_{T,j})/(p_{T,i}+p_{T,j})$. \\
16 & \texttt{max\_dr\_axis} & Larger of the two \texttt{DR\_Axis} values. \\
17 & \texttt{min\_dr\_axis} & Smaller of the two \texttt{DR\_Axis} values. \\
18 & \texttt{max\_isocone} & Larger of the two \texttt{IsoCone} values. \\
19 & \texttt{min\_isocone} & Smaller of the two \texttt{IsoCone} values. \\
20 & \texttt{pair\_pt} & $p_{T,ij}=\sqrt{(p_{x,i}+p_{x,j})^2+(p_{y,i}+p_{y,j})^2}$. \\
21 & \texttt{pair\_eta} & Pseudorapidity of the combined three-momentum $\vec p_i+\vec p_j$. \\
\bottomrule
\end{tabular}
\end{table}

\begin{table}[t]
\centering
\caption{Model-specific pair features used by MLP and ParT, grouped by their physical content. MLP retains separate quantities for the two photons without explicit symmetrization, while the ParT features are symmetric under photon exchange. Both descriptors contain 23 components, but their definitions differ.}
\label{tab:model_pair_features}
\footnotesize
\setlength{\tabcolsep}{5pt}
\renewcommand{\arraystretch}{1.10}
\begin{tabular}{@{}p{0.23\textwidth}c p{0.60\textwidth}@{}}
\toprule
Feature group & Number & Contents \\
\midrule
\multicolumn{3}{c}{\textit{MLP: 23 photon-specific pair features}} \\
\midrule
Diphoton mass & 1 & Invariant mass of the candidate pair. \\
Photon kinematics & 8 & Energy, transverse momentum, pseudorapidity, and azimuth of each photon. \\
Pair separations & 4 & Angular distance, pseudorapidity difference, azimuthal difference, and spatial separation. \\
Kinematic fractions & 4 & Two photon-energy fractions and two transverse-momentum fractions. \\
Axis-distance variables & 3 & The two photon axis distances and their absolute difference. \\
Isolation variables & 3 & The two photon isolation energies and their absolute difference. \\
\midrule
\multicolumn{3}{c}{\textit{ParT: 23 symmetric pair features}} \\
\midrule
Diphoton mass & 1 & Invariant mass of the candidate pair. \\
Energy variables & 4 & Energy sum, absolute energy difference, and the larger and smaller energy fractions. \\
Transverse momenta & 4 & Scalar transverse-momentum sum, absolute difference, and the larger and smaller transverse-momentum fractions. \\
Pseudorapidities & 2 & Mean pseudorapidity and absolute pseudorapidity difference. \\
Angular separations & 3 & $\cos(\Delta\phi)$, $|\Delta\phi|$, and $\Delta R$. \\
Spatial separations & 3 & Three-dimensional separation, transverse separation, and absolute longitudinal separation. \\
Axis-distance variables & 2 & Sum and absolute difference of the two photon axis distances. \\
Isolation variables & 2 & Sum and absolute difference of the two photon isolation energies. \\
Momentum magnitudes & 2 & Sum and absolute difference of the photon momentum magnitudes. \\
\bottomrule
\end{tabular}
\end{table}

\noindent\textbf{Pair-level classifiers.}
BDT and MLP score each candidate from its pair features without encoding the other reconstructed photons as separate objects. The MLP architecture is summarized in table~\ref{tab:network_architectures}, and the BDT configuration is listed in table~\ref{tab:xgboost_training}.

\noindent\textbf{ParticleNet and Particle Transformer.}
For PN and ParT, photon nodes are collected from the unique photon indices in the event's candidate-pair list before the loose mass requirement is applied to individual pairs. The input may therefore include photons that do not form a preselected pair. 
Photon nodes are kept in their original index order, with a mask to distinguish valid nodes from padding.
The maximum number of photon nodes per event is set to 16, 48, 48, and 96 for $Z\to\tau^+\tau^-$, $Z\to q\bar q$, $\nnh$, and $\qqh$, respectively. They are distinct from the neighborhood size $k_{\rm NN}=16$ used in the PN EdgeConv blocks.

PN uses three EdgeConv blocks~\cite{ParticleNet,Wang:2018nkx}. Neighbors in the adopted implementation are determined from a precomputed angular-distance matrix in $(\eta,\phi)$ and retained across the blocks, although the node features are updated. ParT uses six transformer blocks with four attention heads and an embedding width of 192. Its attention logits include a learned bias derived from the symmetric pair features.

For each candidate pair $(i,j)$, the encoded photon representations $h_i$ and $h_j$ are combined as
\begin{equation}
u_{ij}=\bigl[h_i+h_j,\;|h_i-h_j|,\;p_{ij}\bigr],
\label{eq:pair_readout}
\end{equation}
where $p_{ij}$ denotes the model-specific pair features. A classification head maps this descriptor to a candidate score. The symmetric readout has 533 components for PN and 407 for ParT, replacing the global classification readout of the original architectures.

Only reconstructed photons are used as nodes in PN and ParT. The event-axis and isolation variables, however, are computed from the full collection of particle-flow objects (PFOs) and supplied as scalar inputs. MC truth information is never used as an input feature. Because the pair descriptors and architectures both differ, their separate effects cannot be isolated by this comparison.

\begin{table}[t]
\centering
\caption{Photon-level input features used for each reconstructed-photon node
in PN and ParT. The angular distance is defined as
$\Delta R(a,b)=\sqrt{(\eta_a-\eta_b)^2+(\Delta\phi_{ab})^2}$,
with $\Delta\phi$ wrapped to $[-\pi,\pi)$.}
\label{tab:photon_features}
\footnotesize
\setlength{\tabcolsep}{5pt}
\renewcommand{\arraystretch}{1.12}
\begin{tabular}{@{}r p{0.24\textwidth} p{0.62\textwidth}@{}}
\toprule
No. & Feature & Definition \\
\midrule
1 & \texttt{photon\_E}
  & Photon energy $E$, with negative input values clipped to zero. \\

2 & \texttt{photon\_theta}
  & Polar angle $\theta$ obtained from $(p_x,p_y,p_z)$. \\

3 & \texttt{photon\_eta}
  & $\eta=-\ln\!\left[\tan(\theta/2)\right]$. \\

4 & \texttt{photon\_pt}
  & $p_T=E\sin\theta$. \\

5 & \texttt{photon\_log1p\_E}
  & $\ln(1+E)$. \\

6 & \texttt{photon\_cos\_theta}
  & $\cos\theta$. \\

7 & \texttt{photon\_x}
  & Photon position variable $X$. \\

8 & \texttt{photon\_y}
  & Photon position variable $Y$. \\

9 & \texttt{photon\_z}
  & Photon position variable $Z$. \\

10 & \texttt{photon\_dr\_axis}
   & Angular distance $\Delta R(\gamma,\mathrm{axis})$ between the photon and the event axis, defined by the direction of $\vec P_{\mathrm{event}}=\sum_{k\in\mathrm{PFOs}}\vec p_k$. All charged and neutral PFOs are included. The $+z$ direction is used if $\vec P_{\mathrm{event}}=0$. \\

11 & \texttt{photon\_isocone}
   & Energy sum within $\Delta R<0.4$ around the photon, excluding its own energy:
   $\mathrm{IsoCone}(\gamma)= \sum_{k\in\mathrm{PFOs}} E_k\,\mathbf{1}[\Delta R(\gamma,k)<0.4]-E_\gamma$.
   The sum includes charged and neutral PFOs, including the other candidate photon if it lies within the cone. The quantity is expressed in GeV without normalization. \\

\bottomrule
\end{tabular}
\end{table}

Table~\ref{tab:network_architectures} summarizes the network structures. Each PN EdgeConv block applies two $1\times1$ convolution--batch-normalization--ReLU layers to central-node and neighbor-difference features, followed by max aggregation. Its candidate decoder uses ReLU and dropout with probability 0.10. The ParT node embedding consists of two linear layers, each followed by layer normalization, GELU, and dropout. The pair-dependent attention bias is generated by a $23\to64\to4$ network and scaled by 0.1. Self-attention is retained for valid nodes, with zero pair bias on the diagonal.

\begin{table}[t]
\centering
\caption{Network structures used for candidate-pair classification. Widths specify the numbers of hidden or output units. PN and ParT replace a global jet- or event-classification readout with the symmetric pair readout in Eq.~\eqref{eq:pair_readout}.}
\label{tab:network_architectures}
\small
\setlength{\tabcolsep}{4pt}
\renewcommand{\arraystretch}{1.15}
\begin{tabular}{@{}p{0.14\textwidth}p{0.43\textwidth}p{0.35\textwidth}@{}}
\toprule
Model & Encoder & Candidate readout \\
\midrule
MLP &
$23\to256\to128\to64\to1$; ReLU and dropout 0.10 after each hidden linear layer. &
One logit per candidate pair. \\
PN &
Three EdgeConv blocks of widths 64, 128, and 256; two convolutional layers per block; $k_{\rm NN}=16$ and max aggregation. &
Symmetric 533-component descriptor; decoder $533\to256\to128\to1$. \\
ParT &
Node embedding $11\to192\to192$; six pre-norm transformer blocks, four heads, feed-forward width 512, and dropout 0.10. &
Symmetric 407-component descriptor; decoder $407\to256\to128\to1$. \\
\bottomrule
\end{tabular}
\end{table}

\subsection{Training and evaluation}
\label{sec:training}

Training, validation, and test samples are separated at the event level, so that photon pairs from the same event cannot appear in different subsets. At the reference ECAL setting, the validation and final test samples contain 20,000 and 30,000 independent events per process, respectively. All input-normalization parameters are determined from training data. Model selection, score thresholds, and mass-window boundaries are fixed before evaluation on the test samples.

Classifiers are trained separately for each process and ECAL setting, and the mass-window boundaries of the classical baseline are reoptimized at each setting. The resolution scan therefore does not test the transfer of a fixed trained model between detector responses. The resolution and training-size scans report validation-sample performance, whereas the reference-setting tables report performance on independent test events. The resolution scan uses $9.5\times10^5$ training events per setting.

The neural networks minimize binary cross entropy using AdamW. To reduce class imbalance, positive and negative candidates are sampled in a $1:1$ ratio within each training event; when only one class is present, its available candidates are retained. Validation and test samples retain their natural class proportions. For the neural networks, 101 score thresholds uniformly spaced between 0 and 1 are evaluated after each epoch. The checkpoint and threshold giving the highest validation $\metric$ are selected for the final evaluation. The mass-window boundaries of the classical baseline and BDT score thresholds are likewise chosen by maximizing $\metric$ on validation data.

Training-size dependence is studied at $\alpha=0.03$, using samples ranging from approximately 10 to $9.5\times10^5$ training events drawn from a pool of one million events. PN and ParT are trained for 50 epochs, while MLP uses up to 40 epochs with early stopping. The model configurations are kept fixed as the training sample increases. Sample sizes refer to events rather than candidate pairs, whose number varies with the process and mass preselection.

For visualization, we interpolate the sampled performance values as a function of $x=\log_{10}(N_{\mathrm{train}})$, where $N_{\mathrm{train}}$ denotes the number of training events. We use shape-preserving piecewise cubic Hermite interpolation (PCHIP), implemented with \texttt{PchipInterpolator}. The interpolation describes the sampled training-size dependence; no common parametric scaling function is fitted.

Most points in the training-size scan correspond to a single training run. Additional runs with different random seeds were performed for PN (three seeds) and ParT (two seeds) at selected configurations. The PN repetitions include the $Z\to\tau^+\tau^-$ study in figure~\ref{fig:double_s_diagnostic}. These checks give an indication of training stability but do not provide a systematic estimate of training-to-training uncertainty.

All neural models use an initial learning rate of $10^{-3}$ and weight decay of $10^{-4}$. The remaining settings are summarized in table~\ref{tab:neural_training}. MLP features are standardized using training-candidate means and standard deviations, with a standard-deviation floor of $10^{-6}$. PN uses separate normalization statistics for node and pair features. For ParT, node statistics are calculated from valid training photons, while decoder pair features are normalized using candidates entering the training loss. Attention-bias features use separate statistics calculated from valid photon pairs in the training events. Padding is set to zero after normalization. BDT inputs are not standardized.

The BDT settings are listed in table~\ref{tab:xgboost_training}. Training candidates are balanced within each event as for the neural networks. Validation-based early stopping is used with a patience of 100 boosting rounds, after which the score threshold is selected by maximizing validation $\metric$.

Statistical uncertainties at the reference ECAL setting are estimated using event-level bootstrap resampling~\cite{Efron:1979bootstrap}. For all five methods, 20,000 bootstrap replicas are generated from 30,000 independent test events per process. Each replica contains 30,000 events drawn with replacement, with all reconstructed candidates and truth labels of each sampled event kept together. The counts $N_{\rm true\ pair}^{\rm all}$, $N_{\rm true\ pair}$, and $N_{\rm sel\ pair}$ are recalculated for each replica. Resampling at the event level preserves correlations among photon pairs from the same event.

The uncertainties quoted for $\metric$ in table~\ref{tab:breakdown_results} correspond to one standard deviation of the bootstrap distribution, rather than confidence-interval half-widths. Relative gains are calculated from the central values using the classical baseline (MW), not the loose mass preselection. Paired-bootstrap uncertainties of the relative gains are not quoted. Model parameters, score thresholds, and mass-window boundaries remain fixed during resampling. The reported uncertainties therefore account for finite test-sample statistics, but not run-to-run training variations or detector-modeling uncertainties.

\begin{table}[t]
\centering
\caption{Fixed BDT configuration used with XGBoost. Validation is used for early stopping and final score-threshold selection.}
\label{tab:xgboost_training}
\small
\setlength{\tabcolsep}{4pt}
\renewcommand{\arraystretch}{1.10}
\begin{tabular}{@{}p{0.61\columnwidth}p{0.30\columnwidth}@{}}
\toprule
Parameter & Value \\
\midrule
Objective & Binary logistic \\
Tree construction & Histogram \\
Maximum boosting rounds & 2000 \\
Maximum tree depth & 4 \\
Learning rate & 0.03 \\
Row subsampling fraction & 0.8 \\
Column subsampling fraction & 0.8 \\
Minimum child weight & 20 \\
Split penalty $\gamma$ & 0 \\
L1 regularization & 0.1 \\
L2 regularization & 5 \\
Early-stopping patience & 100 rounds \\
Validation metrics & Log loss, AUC \\
\bottomrule
\end{tabular}
\end{table}

\begin{table}[t]
\centering
\caption{Neural-network training and model-selection settings. All models use AdamW with learning rate $10^{-3}$ and weight decay $10^{-4}$. Validation retains the natural candidate-class distribution.}
\label{tab:neural_training}
\small
\setlength{\tabcolsep}{4pt}
\renewcommand{\arraystretch}{1.12}
\begin{tabular}{@{}p{0.26\textwidth}p{0.23\textwidth}p{0.22\textwidth}p{0.22\textwidth}@{}}
\toprule
Setting & MLP & PN & ParT \\
\midrule
Batch size & 8192 candidate pairs & 32 event graphs & 32 event graphs \\
Maximum epochs & 40 & 50 & 50 \\
Early stopping & Patience of 10 epochs & None & None \\
Learning-rate schedule & Cosine, $T_{\max}=40$ & Cosine, $T_{\max}=50$ & Cosine, $T_{\max}=50$ \\
Training loss & Binary cross entropy & Binary cross entropy on sampled edges & Binary cross entropy on sampled edges \\
Class sampling & Per-event $1:1$ & Per-event $1:1$ & Per-event $1:1$ \\
Checkpoint criterion & Highest validation $\metric$ & Highest validation $\metric$ & Highest validation $\metric$ \\
Threshold grid & 101 points in $[0,1]$ & 101 points in $[0,1]$ & 101 points in $[0,1]$ \\
\bottomrule
\end{tabular}
\end{table}

\section{Additional event displays}
\label{sec:additional_event_displays}

\begin{figure}[!tbp]
\centering
\eventdisplay{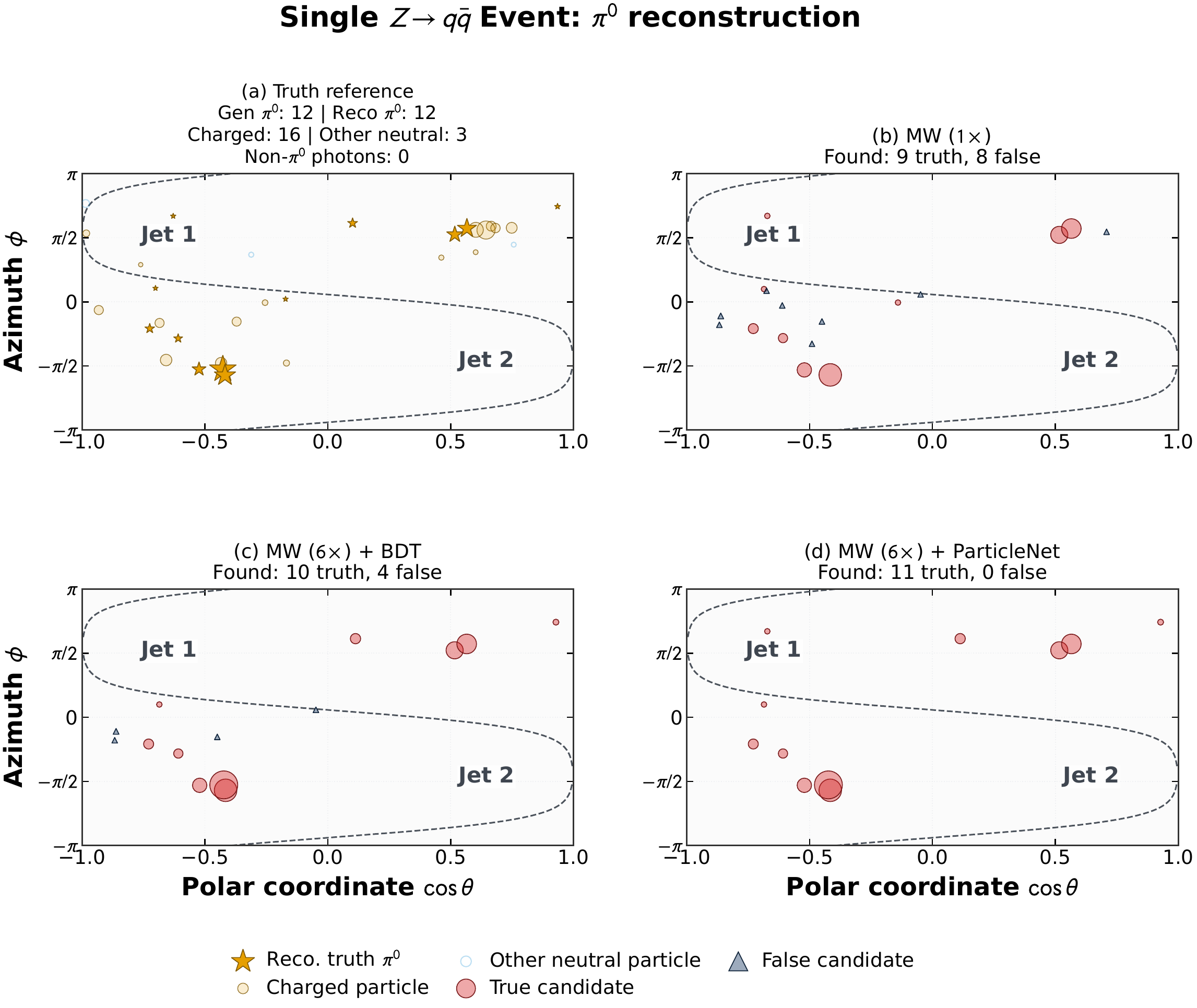}
\caption{An illustrative $Z\to q\bar q$ event at the reference ECAL setting. The panel arrangement and marker definitions follow those in figure~\ref{fig:pn_event_display_Ztautau}, with a $6\times$ mass preselection for BDT and PN.}
\label{fig:pn_event_display_Zqq}
\end{figure}

\begin{figure}[!tbp]
\centering
\eventdisplay{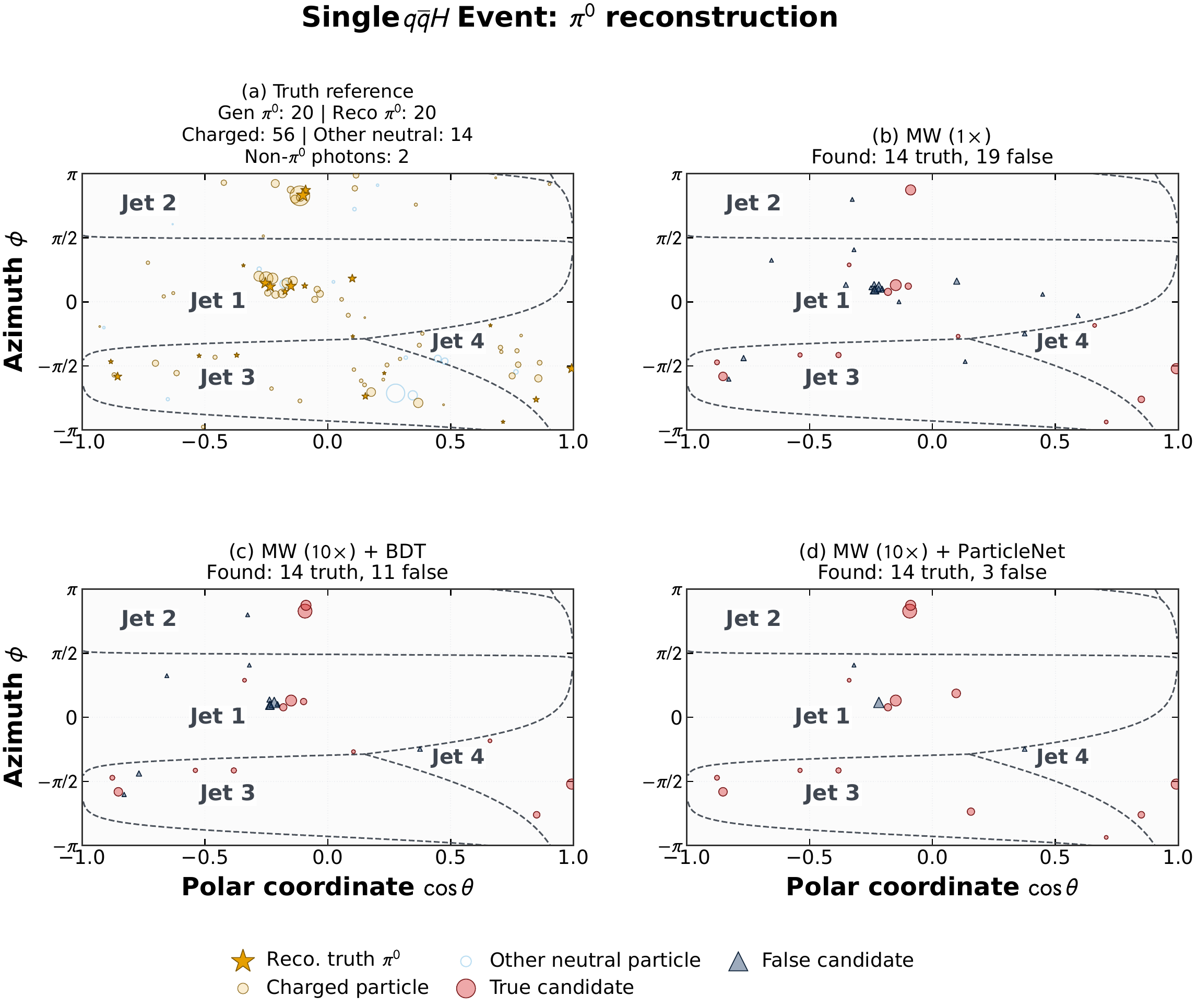}
\caption{An illustrative $\qqh$ event at the reference ECAL setting. The panel arrangement and marker definitions follow those in figure~\ref{fig:pn_event_display_Ztautau}, with a $10\times$ mass preselection for BDT and PN.}
\label{fig:pn_event_display_ZH}
\end{figure}

Figures~\ref{fig:pn_event_display_Zqq} and~\ref{fig:pn_event_display_ZH} show illustrative $Z\to q\bar q$ and $\qqh$ events, complementing the $Z\to\tau^+\tau^-$ and $\nnh$ examples in section~\ref{sec:event_displays}.

\FloatBarrier

\bibliographystyle{JHEP}
\bibliography{references}
\end{document}